\documentclass[journal]{IEEEtran}

\usepackage{graphicx}
\usepackage{amsmath}
\usepackage[usenames, dvipsnames]{color}

\graphicspath{{./Figures/}}

\begin{document}
\title{Practical Demonstration of a Memristive Fuse}

\author{Alexander~Serb,~\IEEEmembership{Member,~IEEE,}
        Ali~Khiat,~\IEEEmembership{}
        and~Themistoklis~Prodromakis,~\IEEEmembership{Senior Member,~IEEE}
\thanks{Alexander Serb, Ali Khiat and Themistoklis Prodromakis are with the Nano Group, ECS, University of Southampton, Highfield, Southampton SO17 1BJ.}
\thanks{Corresponding author email: A.Serb@soton.ac.uk}
\thanks{This work was supported by EPSRC EP/K017829/1 and EU-FP7 RAMP.}}


\maketitle

\begin{abstract}
Since its inception the memristive fuse has been a good example of how small numbers of memristors can be combined to obtain useful behaviours unachievable by individual devices. In this work, we link the memristive fuse concept with that of the Complementary Resistive Switch (CRS), exploit that link to experimentally demonstrate a practical memristive fuse using $TiOx$-based ReRAM cells and explain its basic operational principles. The fuse is stimulated by trains of identical pulses where successive pulse trains feature opposite polarities. In response, we observe a gradual (analogue) drop in resistive state followed by a gradual recovery phase regardless of input stimulus polarity; echoing traditional, binary CRS behaviour. This analogue switching property opens the possibility of operating the memristive fuse as a single-component step change detector. Moreover, we discover that the characteristics of the individual memristors used to demonstrate the memristive fuse concept in this work allow our fuse to be operated in a regime where one of the two constituent devices can be switched largely independently from the other. This property, not present in the traditional CRS, indicates that the inherently analogue memristive fuse architecture may support additional operational flexibility through e.g. allowing finer control over its resistive state.
\end{abstract}


%
\IEEEpeerreviewmaketitle

\section{Introduction}\label{introsec}

The development of a bio-inspired computation paradigm capable of demonstrating practical applications has long been a holy grail of electronics and neuroscience research. This effort has been primarily driven by the immense opportunity to be found in the powerful complementarity that exists between the fast, reliable and precise Von Neumann-based computers and the self-adaptive, massively parallel and fault-tolerant biological systems we see in nature. However, efforts in that direction have so far been hampered by the sheer complexity involved in emulating biological processes \textit{in silico}. Thus far, a number of approaches have tried to attack this problem by harnessing the power of Personal Computers (PCs) \cite{Markram06}, Micro-Processor Units (MPUs) \cite{Furber14}, Field-Programmable Gate Arrays (FPGAs) and Graphics Processing Units (GPUs) \cite{Givon16} as well as bespoke systems exploiting analogue Complementary Metal-Oxide Semiconductor (CMOS) technologies \cite{Qiao15, Schemmel10}. The common problem with these approaches, however, is that they all employ fundamental components and design methodologies originally conceived for Von Neumann-based computation. The resulting power and area costs associated with building bio-inspired systems of any appreciable scale (e.g. see \cite{Stromatias13}) have prompted the evolution of alternative approaches.

\begin{figure}[!t]
\centering
\includegraphics[width=8.7cm]{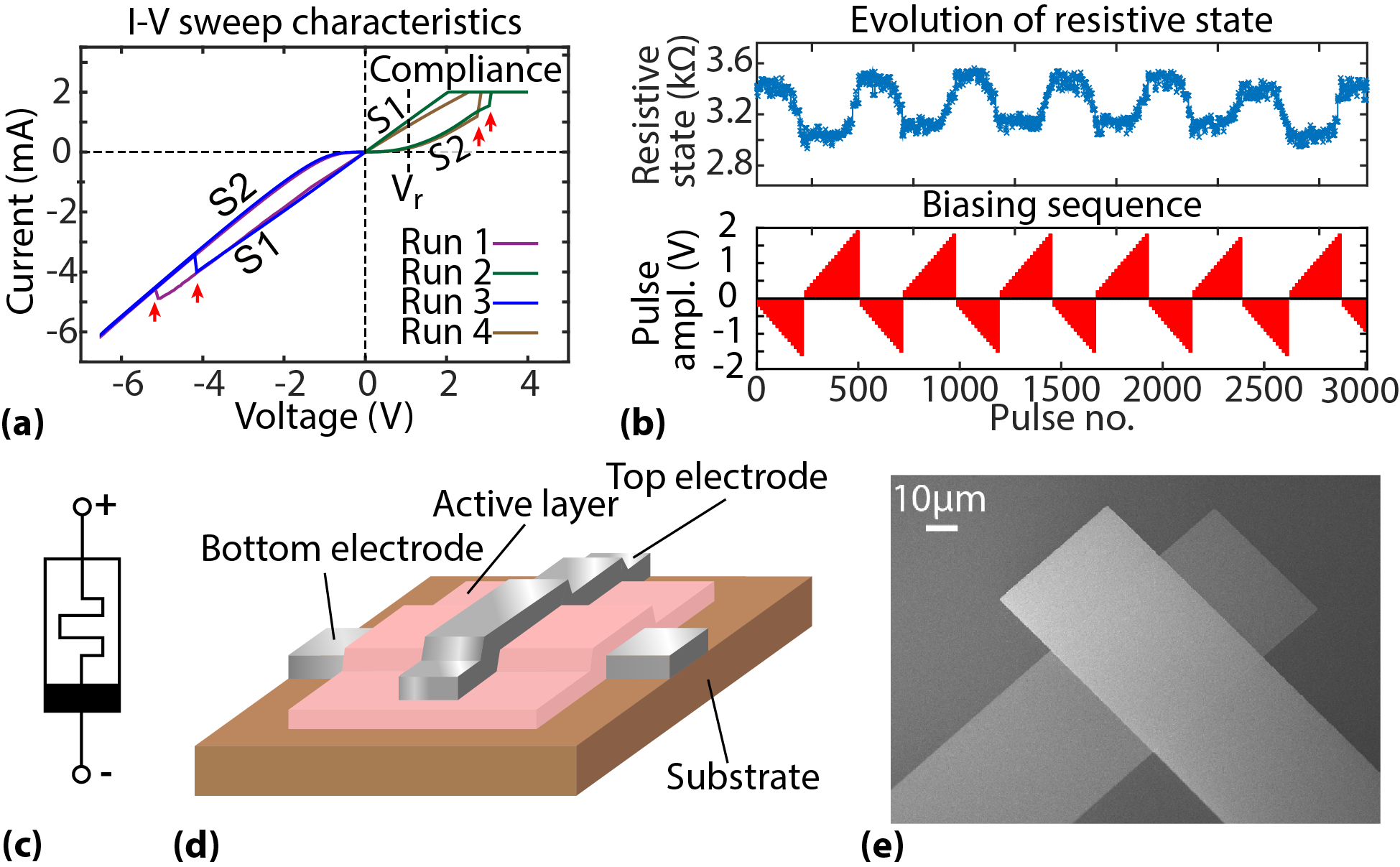}
\caption{Memristors in practice. a) Measured resistive switching in a $TiOx$ ReRAM device. Voltage sweeping causes the memristor to toggle between different resistive states (two such states are marked `S1' and `S2'). In this example toggling occurs abruptly as indicated by the red arrows. The resulting resistive state can then be assessed as static resistance at a standardised read-out voltage $V_r$. In compliance (current-limited) regime, desired voltage values are displayed rather than measured values. b) Measured results showing gradual (analogue) switching under pulse-ramp stimulation for a metal-oxide-based memristor (device M1 in our experiments). c) Memristor circuit symbol. d) Illustration of metal-oxide implementation of a memristor. e) Microphotograph of solid-state metal-oxide-based memristor.}
\label{IntroFig}
\end{figure}

One such approach exploits recent advances in the field of nanoelectronic devices that exhibit the phenomenon of resistive switching \cite{Waser07}, often referred to as `memristors' \cite{Chua71}. Memristors boast a simple, two-terminal structure and an ability to react to external voltage/current stimuli by changing their resistive states \cite{Chua76} as seen in the examples of Fig. \ref{IntroFig}(a,b). These properties, in turn, allow them to act both as single-component memory elements (including synapses \cite{Chiu12, Serrano13, Wei15, Mostafa15}) and as computational elements \cite{Berdan16, Lehtonen09}, which raises the possibility of shifting much neuro-computational complexity to new components designed specifically to exhibit biological neuron- or synapse-like characteristics. These behavioural aspects, in tandem with continuous advances in the electrical and scalability characteristics of memristors \cite{Strachan11, Govoreanu11} indicate a possible route towards truly scalable bio-inspired systems. The flexibility and possibilities offered by this approach are reflected in the variety of memristor implementations and biologically relevant applications investigated so far. These include synapse implementations (analogue synapses supporting Spike Timing-Dependent Plasticity (STDP) implemented through metal-oxide- (MOx) or Phase Change Memory-based (PCM) devices \cite{Jo10, Suri11}, binary stochastic synapses through Spin-Torque Transfer (STT) devices \cite{Vincent15} etc.), small neural networks \cite{Prezioso15}, neural activity sensors \cite{Gupta15} and a host of others, e.g. \cite{Berdan12, Xia09}.

\begin{figure}[!t]
\centering
\includegraphics[width=8.7cm]{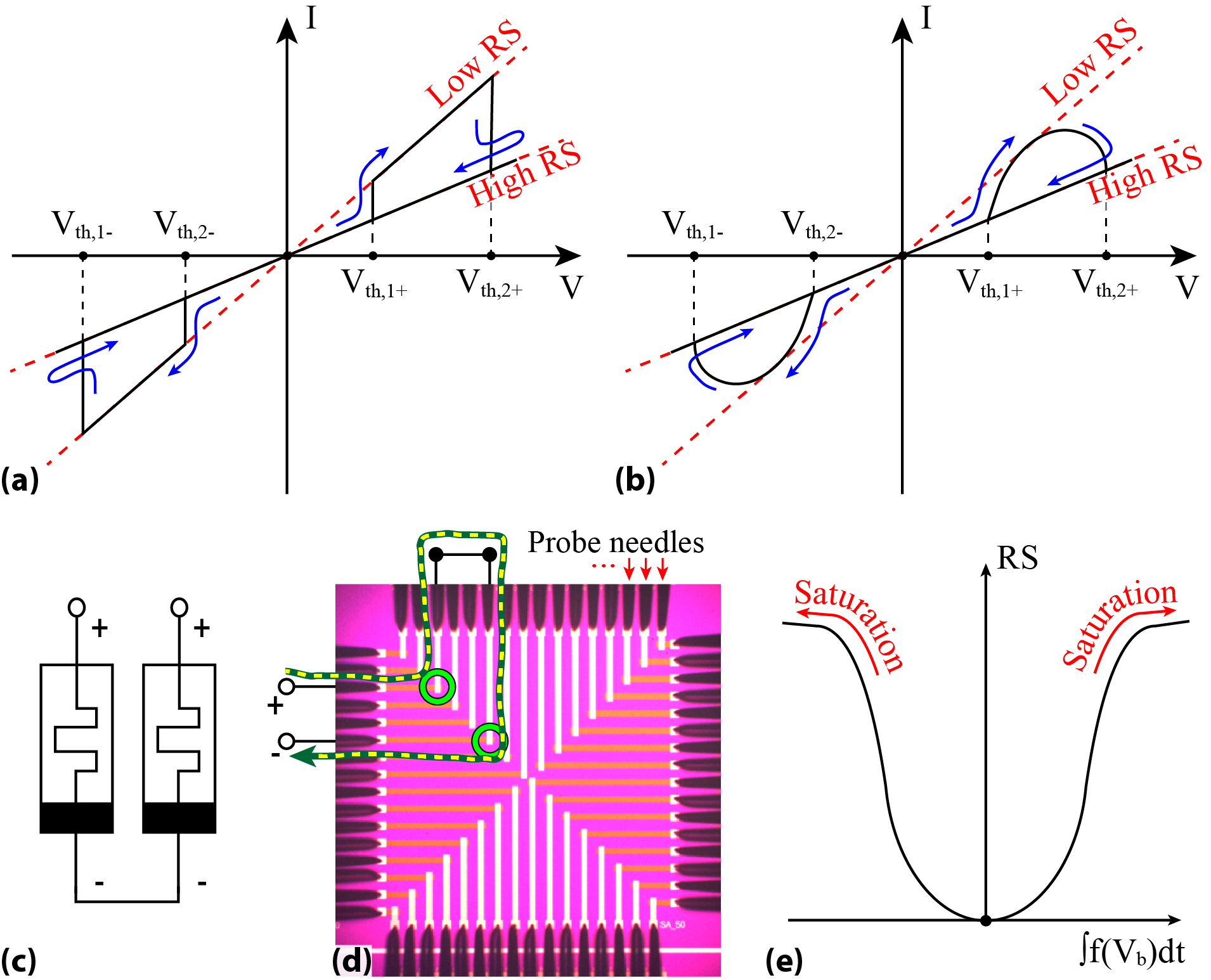}
\caption{Generalising the complementary resistive switch (CRS) concept: a) Conceptual CRS behaviour summary under voltage sweep stimulation. As the voltage is continuously ramped up and down the system transitions abruptly between two distinct resistive states (blue arrows and red dashed lines respectively). b) Expected conceptual behaviour of a memristive fuse under voltage sweeping stimulation. c) Both the CRS and the memristive fuse are implemented as an ensemble of two memristors connected anti-serially. d) Microphotograph of array of memristors in contact with a probe card showing possible connection scheme for linking encircled devices into a fuse configuration. e) Mapping the integral of a transformation of input voltage $V_b$ (through the `switching function', see Fig. \ref{aCRSmechanics}(a)) on an analogue memristive fuse resistive state. Saturation occurs when the memristive fuse reaches its operational resistive state ceiling.}
\label{CRS_aCRS}
\end{figure}

Beyond their use as single-component memory/synapses, memristors can also exhibit interesting properties when operated collectively in small ensembles. One of the most well-known small device ensembles is the complementary resistive switch (CRS) configuration \cite{Linn10} seen in Fig. \ref{CRS_aCRS}(c), i.e. a pair of memristors supporting binary switching (i.e. between two distinct high and low resistive states (HRS, LRS)) connected anti-serially. In typical operation the CRS can only be found in two distinct states: i) an HRS whereby one of the memristors is in HRS and the other in LRS and ii) an LRS when both memristors are in LRS. The CRS changes states abruptly and in a fully voltage level-dependent fashion since each memristor features two threshold levels of opposite polarities at which they carry out their high-low and low-high state transitions, for a total of four CRS thresholds as seen in Fig. \ref{CRS_aCRS}(a). However, if the memristors are allowed to exhibit analogue (gradual) switching at a voltage-dependent rate (Fig. \ref{CRS_aCRS}(b)) the ensemble generalises to a `memristive fuse' as originally described in \cite{Jiang09}. As a result, two new properties may arise: i) The memristive fuse becomes able to assume a large number of resistive state states and ii) the switching no longer occurs consistently abruptly. These properties are important because now the fuse can continuously change states in response to both stimulus duration and voltage amplitude, i.e. it becomes capable of storing information on the history of the input stimulus signal in its overall resistive state (Fig. \ref{CRS_aCRS}(e)).

Such unconventional components that find no direct equivalent in nature may prove useful for complementing the field of bio-inspired computation (`beyond bio-inspired computation'). In this work, we experimentally demonstrate a practical memristive fuse consisting of two metal-oxide-based solid state memristors and show how its intrinsic properties allow it to function as a rudimentary step detector. Section \ref{Methodsec} describes the typical behaviour of individual memristors, presents measured electrical characterisation data from solid-state devices and explains how two exemplars can be combined to result in a memristive fuse. Section \ref{ressec} shows experimental results illustrating fuse behaviour whilst section \ref{Dissec} provides a brief overview of practical fuse operation considerations including avenues for further exploration. Finally, section \ref{sumsec} concludes the paper.


\section{Materials and methods}\label{Methodsec}

\subsection{Memristor device fabrication and testing}\label{Devsec}

The tested memristive devices are based on metal-insulator-metal (MIM) structure fabricated on 200 nm insulating $SiO_2$ film, which was thermally grown on silicon wafer. The bottom and top electrodes (BE, TE) were deposited via electron beam evaporation technique, where the active layer was deposited by reactive magnetron sputtering. All layers were patterned and defined by conventional optical lithography and lift-off processes. Oxygen plasma cleaning step was carried out before each material deposition for obtaining more reliable and better quality devices. The final stack consists of $Pt/TiOx/Pt/Ti$ with respective thicknesses of $10/25/10/5\,nm$. A $Ti$ layer was needed for adhesion purposes and the resulting $TiOx$ film was near stoichiometric. Figures \ref{IntroFig}(d,e) depict a schematic view and SEM micro-photograph of a single memristor cell respectively.

All experiments performed towards this work employed an upgraded version of the memristor characterisation instrument reported in \cite{Berdan15}. All devices involved were probed directly on-wafer via a probe-card as illustrated in Fig. \ref{CRS_aCRS}(d). The BE was always kept grounded and all quoted voltages refer to the TE.
 

\subsection{Memristive fuse basic operation concept}\label{aCRSconc}

\begin{figure}[!t]
\centering
\includegraphics[width=8.5cm]{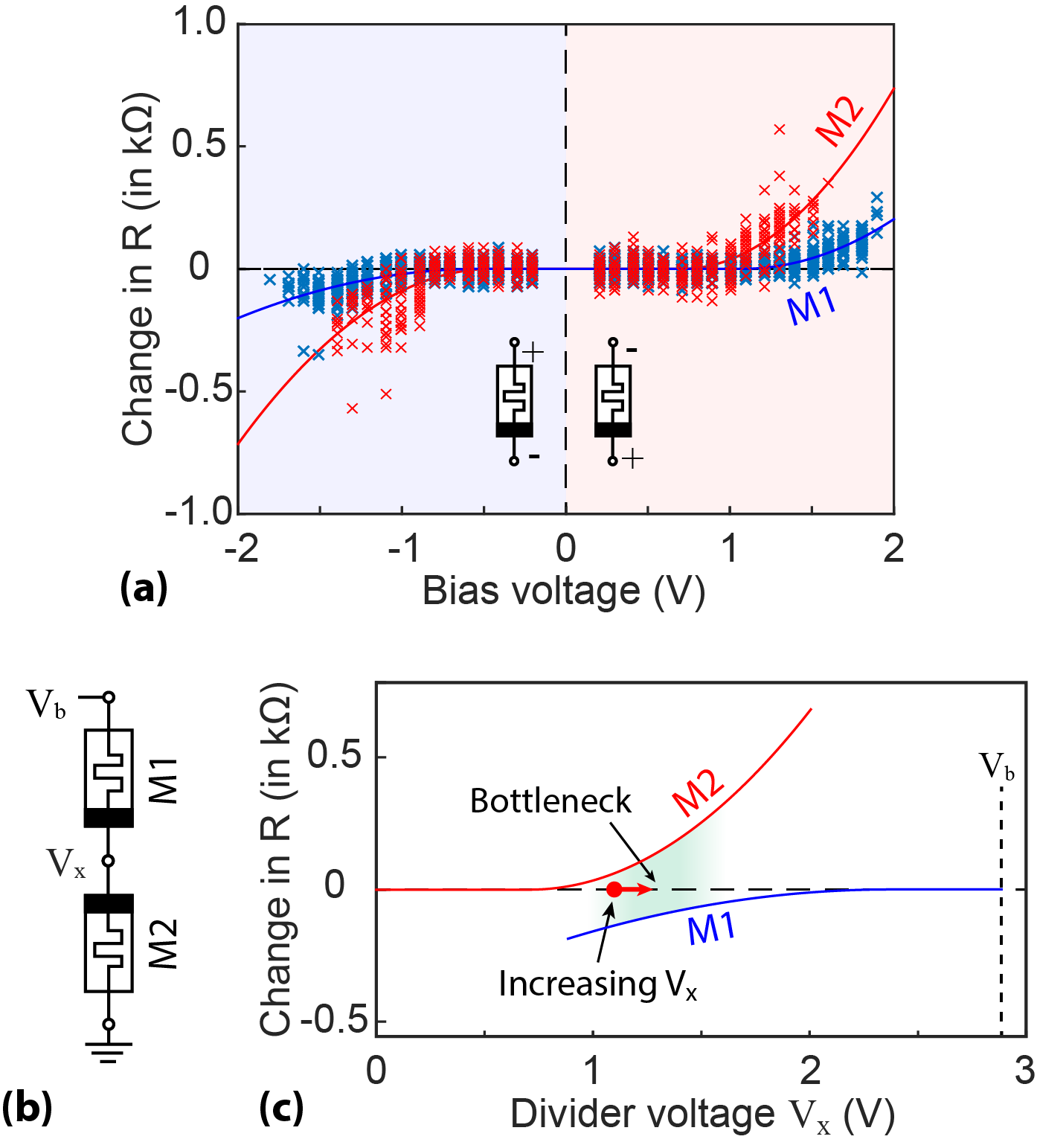}
\caption{Memristive fuse implementation: a) Switching function for the two devices used to construct the memristive fuse and fittings to model in eq. \ref{memodel}. Associated fitting parameters are in table \ref{aCRSfits}. Convention for forward and reverse bias regions (pale blue and red respectively) of memristor operation also shown. b) Connectivity of test devices for memristive fuse experiments in this work. c) Example of a `switching load line' for the two devices from (a) connected as indicated in (b) and with pulse bias voltage $V_b$ slightly lower than $3\,V$.}
\label{aCRSmechanics}
\end{figure}

The operation of the memristive fuse arises naturally from the resistive switching characteristics of its constituent memristors, and specifically the link between input voltage and degree of resistive switching under fixed-duration pulsed stimulation. The sensitivity of switching to input voltage amplitude can be assessed by applying a series of voltage pulse train ramps to each device and measuring the resistive state of the Device Under Test (DUT) at the end of each train as seen in Fig. \ref{IntroFig}(b) and described in detail in \cite{Serb15}. Subsequently the resistive state change precipitated by each voltage level tested is assessed and the relation between resistive state change ($\Delta R$) and applied voltage is summarised in a `voltage sensitivity' plot (or `switching' plot). Measured and fitted switching plots for pulse duration fixed at $100\,\mu s$ are shown in Fig. \ref{aCRSmechanics}(a) for the two devices used to implement the memristive fuse in section \ref{ressec}. The fitting model used is a simple, empirical, four parameter model:

\begin{equation}\label{memodel}
\Delta R(V_b) = \begin{cases} a_+ \cdot (V_b - V_{th+})^2, & V_b > V_{th+} \\ 0, & V_{th-} < V_b < V_{th+} \\ a_- \cdot (V_b - V_{th-})^2, & V_b < V_{th-} \end{cases}
\end{equation}
where $\Delta R(V_b)$ is the change in resistive state, $a_{+,-}$ fitted scaling parameters, $V_{th+,-}$ the fit-estimated thresholds of the memristor and $V_b$ the bias voltage applied across it. Memristor resistive state read-out operations are in all cases carried out at a standardised read-out voltage of $+0.2\,V$ and resistive state is formally defined as static resistance at that voltage level. This is necessary in order to provide a comparable means of assessing resistive state for devices that may feature non-linear I-V characteristics \cite{Regoutz16}.

\begin{table}
\centering
\caption{Fitting (above separator line) and resistive state operating (below separator) parameters for devices used in memristive fuse experiments.}
\label{aCRSfits}
\begin{tabular}{c|ccc}
\textbf{Parameter}          & \textbf{M1}             & \textbf{M2}    &   \textbf{Units}   \\ \hline
$V_{th+}$ & 1.07                      & 0.71                   &    V  \\
$V_{th-}$ & -0.52                     & -0.45                  &   V   \\
$a_+$     & 235.2                       & 439.4                    &    $\Omega/V^2$  \\
$a_-$     & -91.8 			      	& -298.2			    &    $\Omega/V^2$	\\ \hline \hline
Base resistive state  &  $\approx$3.3          & $\approx$4.5         & $k\Omega$ \\
Resistive state range &  3.0 - 3.6                 & 4.1 - 4.9               & $k\Omega$ 
\end{tabular}
\end{table}

The shape of the switching plot strongly determines memristive fuse operation: drawing inspiration from the load line analysis technique we can combine the switching fittings into a `switching load line' for an anti-serial connection of two devices as shown in Fig. \ref{aCRSmechanics}(b). Fig. \ref{aCRSmechanics}(c) shows a load line example for switching under positive bias voltage $V_b$. Because of the anti-serial connection, the switching load line consists of the positive $V_b$ segment of the voltage sensitivity function of M2 and the negative $V_b$ segment for M1. Given some initial resistive state for each device and assuming that the voltage-dependence of the switching rate of each device is relatively independent of its running resistive state (approximately true within a sufficiently small resistive state range) we can use the switching load line to start extracting information on the expected, conceptual behaviour of the memristive fuse.

We begin by noting that pulsing the memristive fuse at $V_b$ will always cause M1 and M2 to experience an decrease/increase in their resistive states respectively and hence their potential divider voltage ($V_x$) increases with each applied pulse, as marked in Fig. \ref{aCRSmechanics}(c). Notably, the precise trajectory of $V_x$ will depend on both the shape of the switching plot and the precise shape of the IV curves of the devices involved, but the effects are always the same: a) the balance of voltage distribution between the two devices in the divider shifts from one device to the other and b) eventually, M2 will saturate at its operational resistive state ceiling (`reset' process) and M1 will saturate at its resistive state floor (`set' process) therefore stabilising the memristive fuse at a relatively high resistive state; similar to the HRS state in the traditional, binary CRS.

Furthermore we observe that the pulse voltage amplitude has been set in such way that the switching load lines form a `bottleneck', i.e. a region in the divider voltage space whereby both memristors experience relatively small changes in resistive state. For appropriate initial values of $V_x$ (as in example of Fig. \ref{aCRSmechanics}(c)) the net effect is that during the early stages of pulsing at constant amplitude $V_b$ the memristive fuse will experience a drop in overall resistance driven by M1 whilst in later stages it will experience a rise in overall resistance driven by M2. This intermediate stage where the fuse features relatively low resistive state is similar to the traditional CRS LRS where both devices in the complementary switch are at their resistive state floors. Memristive fuse operation is similar when the polarity of $V_b$ is reversed. We have thus created a simple, analogue circuit component that encodes the accumulation of many same polarity pulsing events as an HRS whilst reacting to unexpected, anti-polar pulsing events by dropping its resistive state, as shown in Fig. \ref{CRS_aCRS}(e).

\section{Experimental results}\label{ressec}

\begin{figure}[!t]
\centering
\includegraphics[width=8.5cm]{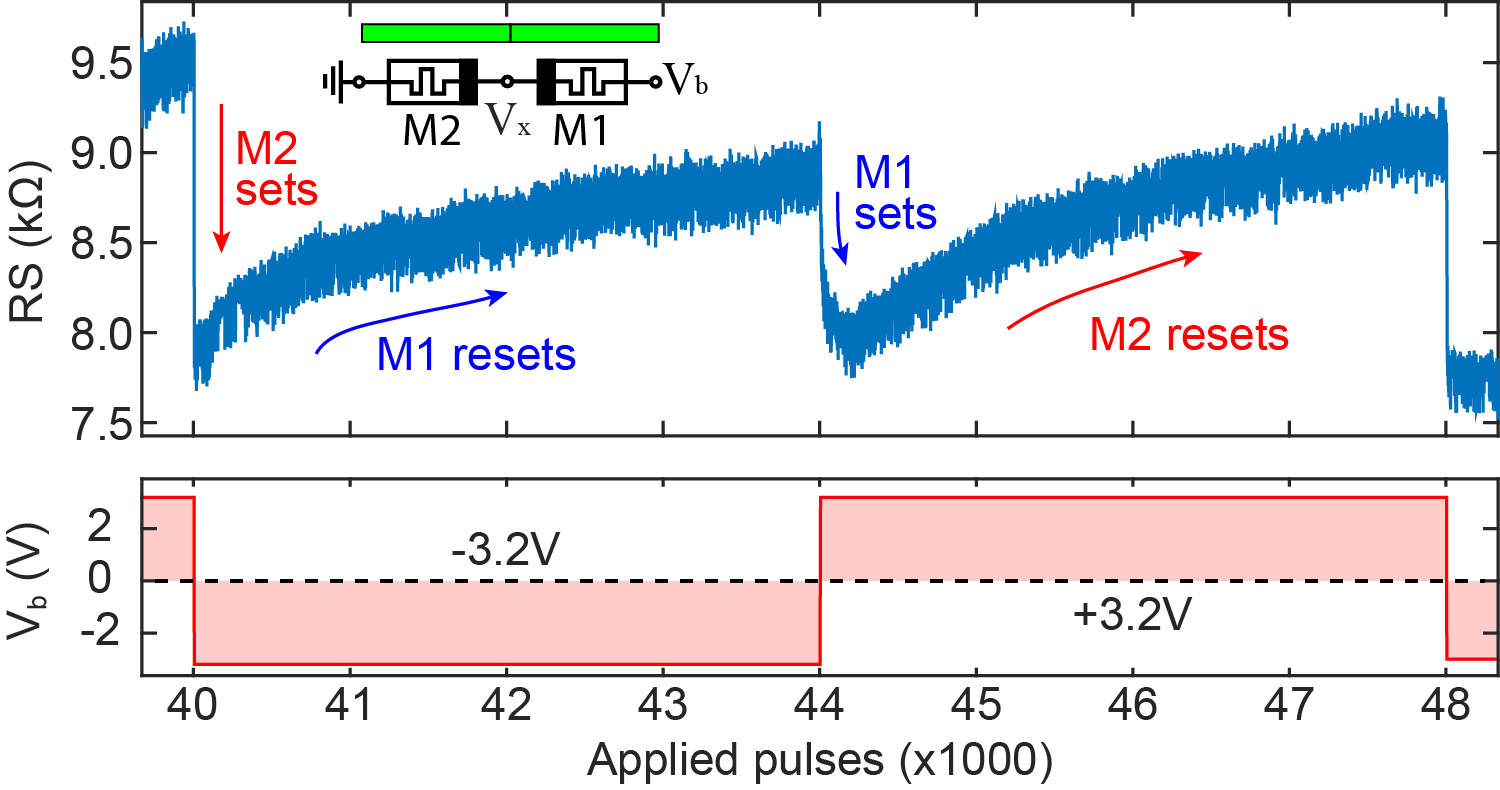}
\caption{Experimental demonstration of memristive fuse behaviour. The resistive state of the fuse (top trace) shows a `dip and recovery' response to both trains of pulses regardless of the polarity of the input signal (bottom trace). Red and blue arrows show which device dominates fuse behaviour as the input pulse sequence progresses. Inset: Under these biasing conditions both memristors operate normally, i.e. above their switching thresholds  (green bars).}
\label{experisults}
\end{figure}

The proposed memristive fuse topology was tested experimentally using the devices from Fig. \ref{aCRSmechanics}(a) connected anti-serially as illustrated in Fig. \ref{CRS_aCRS}. Results are shown in Fig. \ref{experisults}. Following initialisation to a saturated state the memrsitive fuse reacts to two trains of pulses with opposite polarities and suitably chosen amplitudes in a qualitatively similar manner: exhibiting a sharp, initial `dip' followed by a slower `recovery' phase as projected.

We notice that the two `dip and recovery' responses are subtly different. With negative stimulus polarity the dip is sharper and the recovery slower than in the positive stimulus case. Closer inspection suggests this may be linked to the balance between the sensitivities of each device to voltage and the resistive state ratio of the devices affecting the precise distribution of voltage within the divider. M2 is slightly more sensitive to voltage than M1, as shown in Fig. \ref{aCRSmechanics}(a) whilst at the same time it operates at a higher resistive state than M1 and hence should capture more of the voltage applied across the memristive fuse provided the IV curves for both devices are qualitatively similar. As a result, the overall fuse response is slower when M1 dominates behaviour.

\begin{figure}[!t]
\centering
\includegraphics[width=8.5cm]{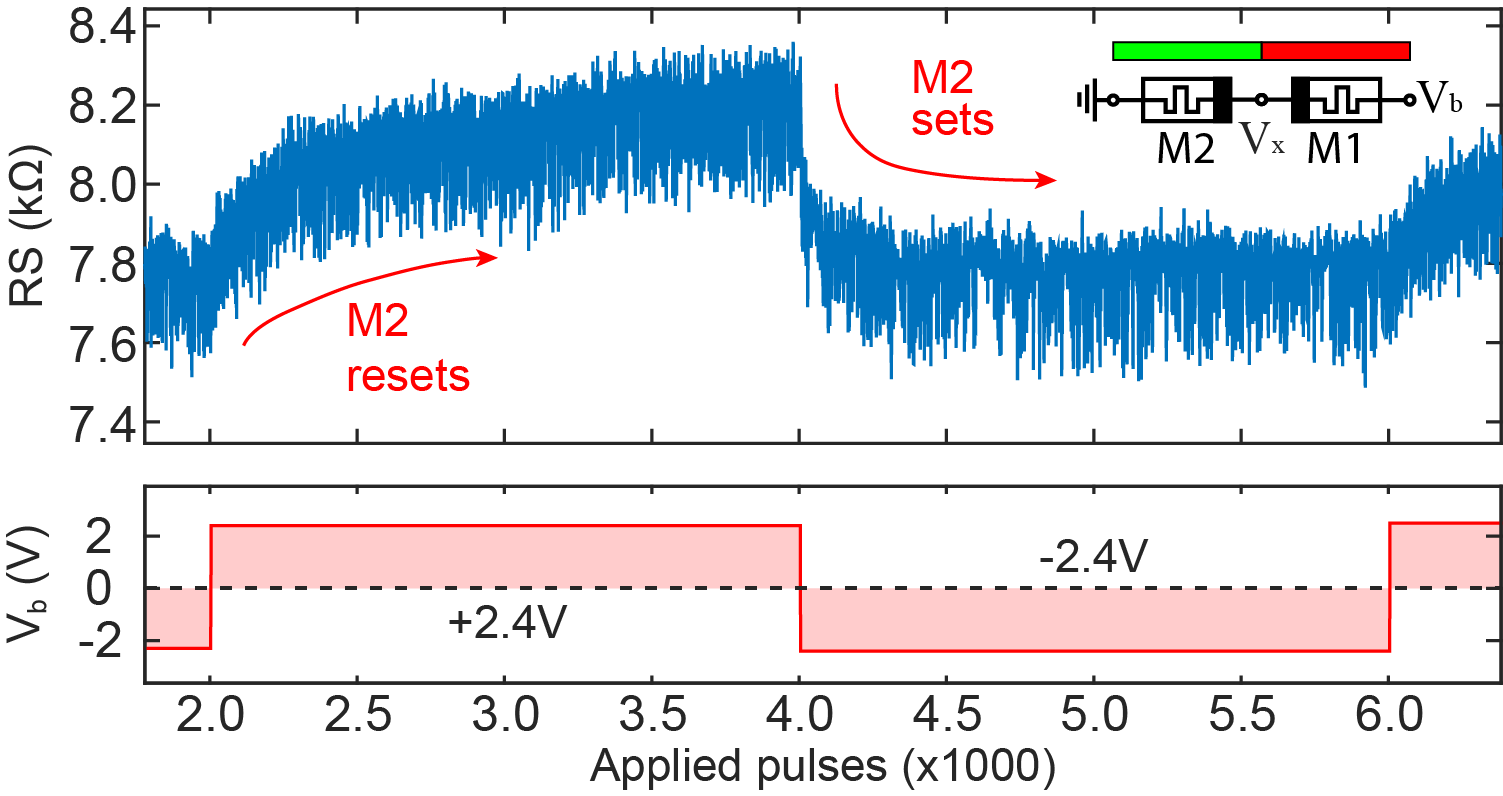}
\caption{Exploiting device mismatch to expand memristive fuse capabilities. The resistive state of the fuse is controlled through exclusive manipulation of M2 (top trace) when the input stimulus voltage (bottom trace) is strong enough to influence M2, but not M1 (inset). Inset: Only device M2 influences switching in this biasing regime (green bar) whilst M1 seems to remain largely inert (red bar).}
\label{experisults2}
\end{figure}

The mismatch in the voltage sensitivity of the two devices in the memristive fuse has a further implication: if the fuse is operated at lower voltages it might be possible to exert an influence on the resistive state of M2 only whilst leaving M1 largely unaffected. Experimental results in Fig. \ref{experisults2} show that this is indeed the case. The memristive fuse sets and resets in a completely bipolar fashion (opposite polarity has opposite effect on resistive state) fully consistent with the voltage sensitivity plot for M2, although at higher voltages in comparison to solo operation of M2. Whether this isolation of M2 can be achieved throughout the entire resistive state range of M1 and if so under what specific biasing circumstances requires further, dedicated study. The ability to selectively exert control on only one of the two devices in the memristive fuse may allow access to far more flexible modes of fuse operation, for example opportunities to set fuse resistive state to an `ultra-HRS' level where both constituent memristors at their operational resistive state ceilings; a situation normally inaccessible in the traditional CRS topology. The precise interrelation between operating voltages and memrsitive fuse operating regimes is a complex topic that merits further, dedicated study.

\section{Discussion}\label{Dissec}

\begin{figure}[!t]
\centering
\includegraphics[width=8.5cm]{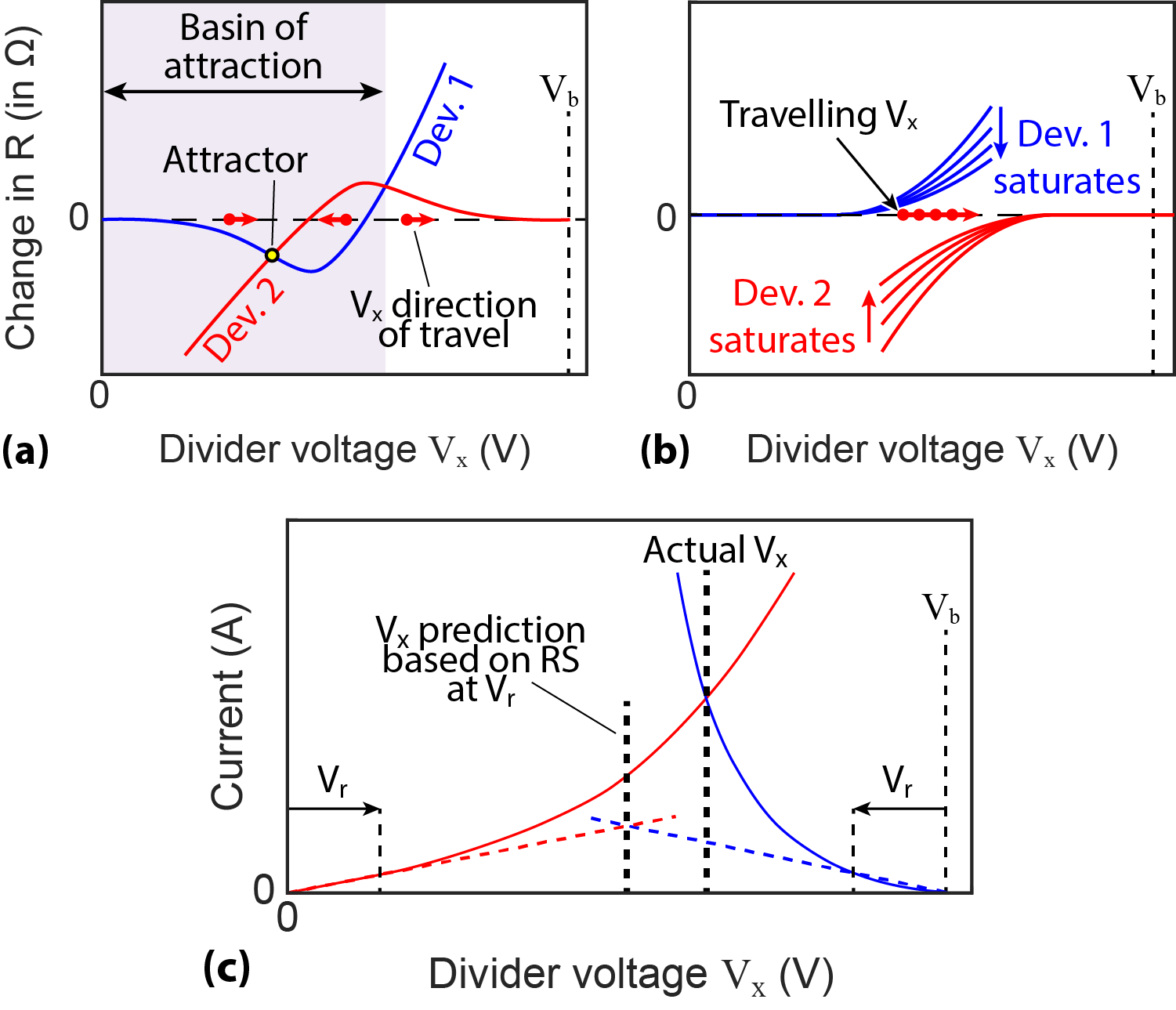}
\caption{Understanding the complexity of the memristive fuse behaviour. a) If switching functions are not monotonic in $V_x$ then the fuse can potentially react to stimuli of appropriate amplitude $V_b$ by converging their resistive states towards/away from specific attractors/repellers. b) The dependence of switching voltage sensitivity to the running resistive state of each constituent device of the fuse implies that as the fuse changes state the switching load line and consequently the shape of the `bottleneck' area is constantly being reshaped. c) Non-linearities and asymmetries in device I-V render predicting the constantly shifting divider voltage $V_x$ for given bias voltage $V_b$ in operando, and consequently the precise degree of switching, difficult.}
\label{discusspts}
\end{figure}

In this work we used a simplified description of memristor operation in section \ref{Methodsec} in order to offer a basic, concept-level explanation of the observed fuse ensemble functionality. However, practical memristive devices typically exhibit rich dynamics far beyond what our `well-behaved' switching plots can capture. Let us review our assumptions and consider the implications when they no longer hold:

`The switching plot can be modelled by a monotonic function of bias voltage': In practice it has been observed that devices can exhibit non-monotonic switching plots like the example in \cite{Serb15} (Fig. 4(b2)) where the switching plot exhibits a curvature reminiscent of $f(x) = a \cdot x^3$. Such switching characteristics would imply that the switching load-lines in Fig. \ref{aCRSmechanics}(c) might cross for appropriately selected bias voltages and thus automatically define fixed points that the memristive fuse could be forced to converge to (attractors) if initialised within the corresponding basin of attraction (Fig. \ref{discusspts}(a)). Nevertheless, so long as the memristive fuse constituent devices are operated at a voltage where the switching load lines do not cross, the fundamental behaviours seen in section \ref{ressec} are in principle preserved.

`The switching function is independent of running resistive state': In practice this dependence can be very complex, but in the simple case where running resistive state influences switchability in an approximately multiplicative way we can describe the effect through a window function which `stretches' the switching function (and therefore also the switching load lines from Fig. \ref{aCRSmechanics}(b)) in the y-direction (Fig. \ref{discusspts}(b)). In that situation, the precise opening of the bottleneck region will continuously change during operation as the memristors change their resistive states, but the qualitative behaviour from section \ref{ressec} will still be preserved so long as the bottleneck doesn't completely disappear (i.e. switching load lines do not cross). The precise effects of this complexity require further, dedicated study.


`Memristor I-V is linear': Non-linearities and asymmetries in the I-V characteristics of the memristors will affect the precise distribution of voltage between the two halves of the memristive fuse (Fig. \ref{discusspts}(c)). This breaks any direct link between measured individual memristor resistive state (carried out at a standard read-out voltage) and actual potential divider voltage under fuse pulse stimulation conditions. For the same reason the resistive state of the entire fuse acting as a single component is different to the sum of the as-measured individual memristor component resistive states. However, so long as a change in `as measured' $\frac{R(M_1)}{R(M_2)}$ ratio translates into a shift of the potential divider voltage in the expected direction the qualitative behaviour of the memristive fuse will be preserved. The precise conditions under which this shall occur are a subject of further study.

The memrsitive fuse's intrinsic properties hint towards some interesting applications: The observed relationship between dip and recovery indicates an inherent ability of the fuse to respond quickly and strongly when a series of many pulses of the same polarity is suddenly interrupted by stimuli of the opposite polarity, i.e. to detect sudden changes in input signal polarity regardless of the actual polarities of all stimuli involved. This opens up the possibility for the application of the memristive fuse as a rudimentary, single-component step detection element for bio-inspired computation. When operated in such manner long series of same polarity events in the input data-stream will be encoded into high fuse resistive state values whilst the arrival of even relatively small numbers of `novel' signals will appear as substantial drops in fuse resistive state. In biological terms, the memristive fuse will therefore act as a `direction of change-independent' variation of the classical adaptable neuron\footnote{Which responds to sudden increases in input stimulation with a transient increase in its firing rate.}. Notably, higher level applications exploiting this behaviour have already been proposed where memristors are connected in square \cite{Jiang09} or hexagonal \cite{Gelencser12} pixel grids,  i.e. connectivity patterns similar to those observed in the outer plexiform layer of the retina \cite{Kolb70}, for the purpose of image edge detection in a biomimetic fashion.

\section{Summary}\label{sumsec}

Overall, in this work we have: a) presented an analogue generalisation of the well-known CRS concept, b) provided a simple and intuitive link between basic device characteristics and expected memristive fuse operation, c) showed experimental evidence of fuse behaviour in metal-oxide memristor pair ensembles and d) made some important observations regarding the expected influence of three key memristor properties on the precise fuse characteristics (voltage sensitivity of switching, resistive state-dependence of switching, I-V non-linearity). We have also concluded that whilst they will undoubtedly affect behaviour quantitatively, the qualitative aspects of fuse behaviour are expected to be conserved so long as a few simple but fundamental assumptions hold. Our discussion highlighted the emergence of a wealth of complexities as the traditional, binary CRS is generalised to a memrsitive fuse; complexities that must either be mitigated to allow good single-component step detector operation or engineered to allow the memristive fuse to fulfil entirely different functions in a single component (both currently under investigation). Finally, we have offered a glimpse into how the inherent properties of the proposed memristive fuse may allow it to find interesting applications, exemplifying its ability to act as a simple, two-terminal, single-component step detector.

\ifCLASSOPTIONcaptionsoff
  \newpage
\fi

\bibliographystyle{IEEEtran}
\bibliography{library}

\begin{thebibliography}{10}
\providecommand{\url}[1]{#1}
\csname url@samestyle\endcsname
\providecommand{\newblock}{\relax}
\providecommand{\bibinfo}[2]{#2}
\providecommand{\BIBentrySTDinterwordspacing}{\spaceskip=0pt\relax}
\providecommand{\BIBentryALTinterwordstretchfactor}{4}
\providecommand{\BIBentryALTinterwordspacing}{\spaceskip=\fontdimen2\font plus
\BIBentryALTinterwordstretchfactor\fontdimen3\font minus
  \fontdimen4\font\relax}
\providecommand{\BIBforeignlanguage}[2]{{%
\expandafter\ifx\csname l@#1\endcsname\relax
\typeout{** WARNING: IEEEtran.bst: No hyphenation pattern has been}%
\typeout{** loaded for the language `#1'. Using the pattern for}%
\typeout{** the default language instead.}%
\else
\language=\csname l@#1\endcsname
\fi
#2}}
\providecommand{\BIBdecl}{\relax}
\BIBdecl

\bibitem{Markram06}
H.~Markram, ``The blue brain project,'' \emph{Nature Reviews Neuroscience},
  vol.~7, no.~2, pp. 153--160, 2006.

\bibitem{Furber14}
S.~B. Furber, F.~Galluppi, S.~Temple, and L.~A. Plana, ``The spinnaker
  project,'' \emph{Proceedings of the IEEE}, vol. 102, no.~5, pp. 652--665,
  2014.

\bibitem{Givon16}
L.~E. Givon and A.~A. Lazar, ``Neurokernel: An open source platform for
  emulating the fruit fly brain,'' \emph{PloS one}, vol.~11, no.~1, 2016.

\bibitem{Qiao15}
N.~Qiao, H.~Mostafa, F.~Corradi, M.~Osswald, F.~Stefanini, D.~Sumislawska, and
  G.~Indiveri, ``A reconfigurable on-line learning spiking neuromorphic
  processor comprising 256 neurons and 128k synapses,'' \emph{Frontiers in
  neuroscience}, vol.~9, p. 141, 2015.

\bibitem{Schemmel10}
J.~Schemmel, D.~Briiderle, A.~Griibl, M.~Hock, K.~Meier, and S.~Millner, ``A
  wafer-scale neuromorphic hardware system for large-scale neural modeling,''
  in \emph{Proceedings of 2010 IEEE International Symposium on Circuits and
  Systems}, May 2010, pp. 1947--1950.

\bibitem{Stromatias13}
E.~Stromatias, F.~Galluppi, C.~Patterson, and S.~Furber, ``Power analysis of
  large-scale, real-time neural networks on spinnaker,'' in \emph{Neural
  Networks (IJCNN), The 2013 International Joint Conference on}, Aug 2013, pp.
  1--8.

\bibitem{Waser07}
R.~Waser and M.~Aono, ``Nanoionics-based resistive switching memories,''
  \emph{Nature materials}, vol.~6, no.~11, pp. 833--840, 2007.

\bibitem{Chua71}
L.~Chua, ``Memristor-the missing circuit element,'' \emph{IEEE Transactions on
  circuit theory}, vol.~18, no.~5, pp. 507--519, 1971.

\bibitem{Chua76}
L.~O. Chua and S.~M. Kang, ``Memristive devices and systems,''
  \emph{Proceedings of the IEEE}, vol.~64, no.~2, pp. 209--223, 1976.

\bibitem{Chiu12}
P.-F. Chiu, M.-F. Chang, C.-W. Wu, C.-H. Chuang, S.-S. Sheu, Y.-S. Chen, and
  M.-J. Tsai, ``Low store energy, low vddmin, 8t2r nonvolatile latch and sram
  with vertical-stacked resistive memory (memristor) devices for low power
  mobile applications,'' \emph{Solid-State Circuits, IEEE Journal of}, vol.~47,
  no.~6, pp. 1483--1496, 2012.

\bibitem{Serrano13}
T.~Serrano-Gotarredona, T.~Masquelier, T.~Prodromakis, G.~Indiveri, and
  B.~Linares-Barranco, ``Stdp and stdp variations with memristors for spiking
  neuromorphic learning systems,'' 2013.

\bibitem{Wei15}
S.~L. Wei, E.~Vasilaki, A.~Khiat, I.~Salaoru, R.~Berdan, and T.~Prodromakis,
  ``Emulating long-term synaptic dynamics with memristive devices,''
  \emph{arXiv preprint arXiv:1509.01998}, 2015.

\bibitem{Mostafa15}
H.~Mostafa, A.~Khiat, A.~Serb, C.~G. Mayr, G.~Indiveri, and T.~Prodromakis,
  ``Implementation of a spike-based perceptron learning rule using tio2- x
  memristors,'' \emph{Frontiers in neuroscience}, vol.~9, 2015.

\bibitem{Berdan16}
R.~Berdan, E.~Vasilaki, A.~Khiat, G.~Indiveri, A.~Serb, and T.~Prodromakis,
  ``Emulating short-term synaptic dynamics with memristive devices,''
  \emph{Scientific reports}, vol.~6, 2016.

\bibitem{Lehtonen09}
E.~Lehtonen and M.~Laiho, ``Stateful implication logic with memristors,'' in
  \emph{Proceedings of the 2009 IEEE/ACM International Symposium on Nanoscale
  Architectures}.\hskip 1em plus 0.5em minus 0.4em\relax IEEE Computer Society,
  2009, pp. 33--36.

\bibitem{Strachan11}
J.~P. Strachan, A.~C. Torrezan, G.~Medeiros-Ribeiro, and R.~S. Williams,
  ``Measuring the switching dynamics and energy efficiency of tantalum oxide
  memristors,'' \emph{Nanotechnology}, vol.~22, no.~50, p. 505402, 2011.

\bibitem{Govoreanu11}
B.~Govoreanu, G.~Kar, Y.~Chen, V.~Paraschiv, S.~Kubicek, A.~Fantini, I.~Radu,
  L.~Goux, S.~Clima, R.~Degraeve \emph{et~al.}, ``10$\times$ 10nm 2 hf/hfo x
  crossbar resistive ram with excellent performance, reliability and low-energy
  operation,'' in \emph{Electron Devices Meeting (IEDM), 2011 IEEE
  International}.\hskip 1em plus 0.5em minus 0.4em\relax IEEE, 2011, pp. 31--6.

\bibitem{Jo10}
S.~H. Jo, T.~Chang, I.~Ebong, B.~B. Bhadviya, P.~Mazumder, and W.~Lu,
  ``Nanoscale memristor device as synapse in neuromorphic systems,'' \emph{Nano
  letters}, vol.~10, no.~4, pp. 1297--1301, 2010.

\bibitem{Suri11}
M.~Suri, O.~Bichler, D.~Querlioz, O.~Cueto, L.~Perniola, V.~Sousa,
  D.~Vuillaume, C.~Gamrat, and B.~DeSalvo, ``Phase change memory as synapse for
  ultra-dense neuromorphic systems: Application to complex visual pattern
  extraction,'' in \emph{Electron Devices Meeting (IEDM), 2011 IEEE
  International}.\hskip 1em plus 0.5em minus 0.4em\relax IEEE, 2011, pp. 4--4.

\bibitem{Vincent15}
A.~F. Vincent, J.~Larroque, N.~Locatelli, N.~Ben~Romdhane, O.~Bichler,
  C.~Gamrat, W.~S. Zhao, J.-O. Klein, S.~Galdin-Retailleau, and D.~Querlioz,
  ``Spin-transfer torque magnetic memory as a stochastic memristive synapse for
  neuromorphic systems,'' \emph{Biomedical Circuits and Systems, IEEE
  Transactions on}, vol.~9, no.~2, pp. 166--174, 2015.

\bibitem{Prezioso15}
M.~Prezioso, F.~Merrikh-Bayat, B.~Hoskins, G.~Adam, K.~K. Likharev, and D.~B.
  Strukov, ``Training and operation of an integrated neuromorphic network based
  on metal-oxide memristors,'' \emph{Nature}, vol. 521, no. 7550, pp. 61--64,
  2015.

\bibitem{Gupta15}
I.~Gupta, A.~Serb, A.~Khiat, R.~Zeitler, S.~Vassanelli, and T.~Prodromakis,
  ``Memristive integrative sensors for neuronal activity,'' \emph{arXiv
  preprint arXiv:1507.06832}, 2015.

\bibitem{Berdan12}
R.~Berdan, T.~Prodromakis, I.~Salaoru, A.~Khiat, and C.~Toumazou, ``Memristive
  devices as parameter setting elements in programmable gain amplifiers,''
  \emph{Applied Physics Letters}, vol. 101, no.~24, p. 243502, 2012.

\bibitem{Xia09}
Q.~Xia, W.~Robinett, M.~W. Cumbie, N.~Banerjee, T.~J. Cardinali, J.~J. Yang,
  W.~Wu, X.~Li, W.~M. Tong, D.~B. Strukov \emph{et~al.}, ``Memristor- cmos
  hybrid integrated circuits for reconfigurable logic,'' \emph{Nano letters},
  vol.~9, no.~10, pp. 3640--3645, 2009.

\bibitem{Linn10}
E.~Linn, R.~Rosezin, C.~K{\"u}geler, and R.~Waser, ``Complementary resistive
  switches for passive nanocrossbar memories,'' \emph{Nature materials},
  vol.~9, no.~5, pp. 403--406, 2010.

\bibitem{Jiang09}
F.~Jiang and B.~E. Shi, ``The memristive grid outperforms the resistive grid
  for edge preserving smoothing,'' in \emph{Circuit Theory and Design, 2009.
  ECCTD 2009. European Conference on}.\hskip 1em plus 0.5em minus 0.4em\relax
  IEEE, 2009, pp. 181--184.

\bibitem{Berdan15}
R.~Berdan, A.~Serb, A.~Khiat, A.~Regoutz, C.~Papavassiliou, and T.~Prodromakis,
  ``A-$\mu$controller-based system for interfacing selectorless rram crossbar
  arrays,'' \emph{Electron Devices, IEEE Transactions on}, vol.~62, no.~7, pp.
  2190--2196, 2015.

\bibitem{Serb15}
A.~Serb, A.~Khiat, and T.~Prodromakis, ``An \mbox{RRAM} biasing parameter
  optimizer,'' \emph{Electron Devices, IEEE Transactions on}, vol.~62, no.~11,
  pp. 3685--3691, 2015.

\bibitem{Regoutz16}
A.~Regoutz, I.~Gupta, A.~Serb, A.~Khiat, F.~Borgatti, T.-L. Lee, C.~Schlueter,
  P.~Torelli, B.~Gobaut, M.~Light \emph{et~al.}, ``Role and optimization of the
  active oxide layer in tio2-based rram,'' \emph{Advanced Functional
  Materials}, vol.~26, no.~4, pp. 507--513, 2016.

\bibitem{Gelencser12}
A.~Gelencser, T.~Prodromakis, C.~Toumazou, and T.~Roska, ``Biomimetic model of
  the outer plexiform layer by incorporating memristive devices,''
  \emph{Physical Review E}, vol.~85, no.~4, p. 041918, 2012.

\bibitem{Kolb70}
H.~Kolb, ``Organization of the outer plexiform layer of the primate retina:
  electron microscopy of golgi-impregnated cells,'' \emph{Philosophical
  Transactions of the Royal Society of London B: Biological Sciences}, vol.
  258, no. 823, pp. 261--283, 1970.

\end{thebibliography}


\vspace*{-3\baselineskip}
\begin{IEEEbiography}[{\includegraphics[width=1in,height=1.25in,clip,keepaspectratio]{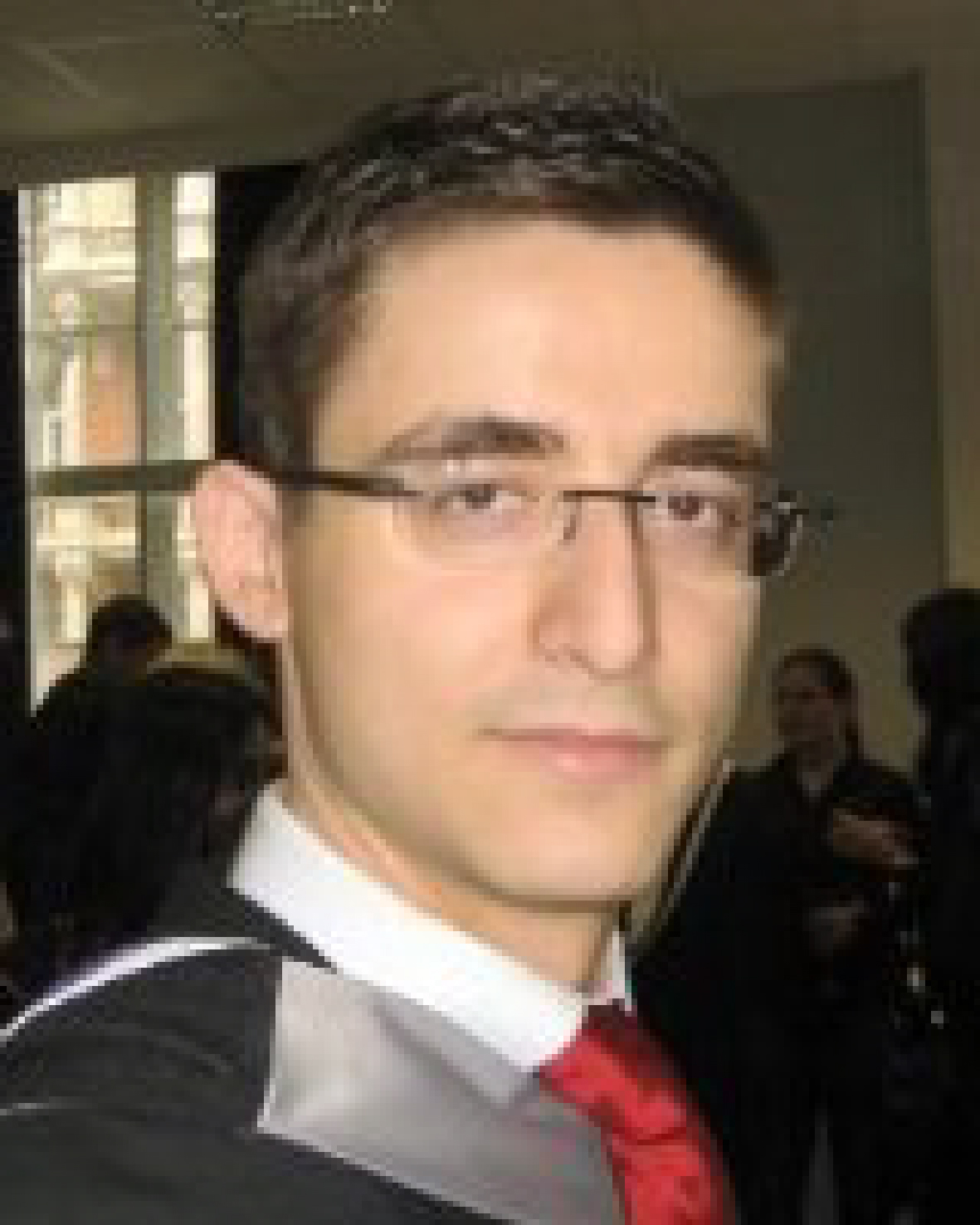}}]{Alexander Serb}
(M'11) is a research fellow at the Electronics and Computer Science (ECS) dept., University of Southampton, UK. His research interests are: instrumentation, algorithms and applications for RRAM testing, and neuro-inspired engineering.
\end{IEEEbiography}

\vspace*{-3\baselineskip}
\begin{IEEEbiography}[{\includegraphics[width=1in,height=1.25in,clip,keepaspectratio]{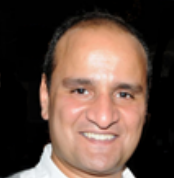}}]{Ali Khiat}
 is an Experimental Officer at Southampton Nanofabrication Centre, University of Southampton. His current main research interests are micro-/nano-fabrication, optimisation, metrology and characterization of memristors and memristive devices.
\end{IEEEbiography}

\vspace*{-3\baselineskip}
\begin{IEEEbiography}[{\includegraphics[width=1in,height=1.25in,clip,keepaspectratio]{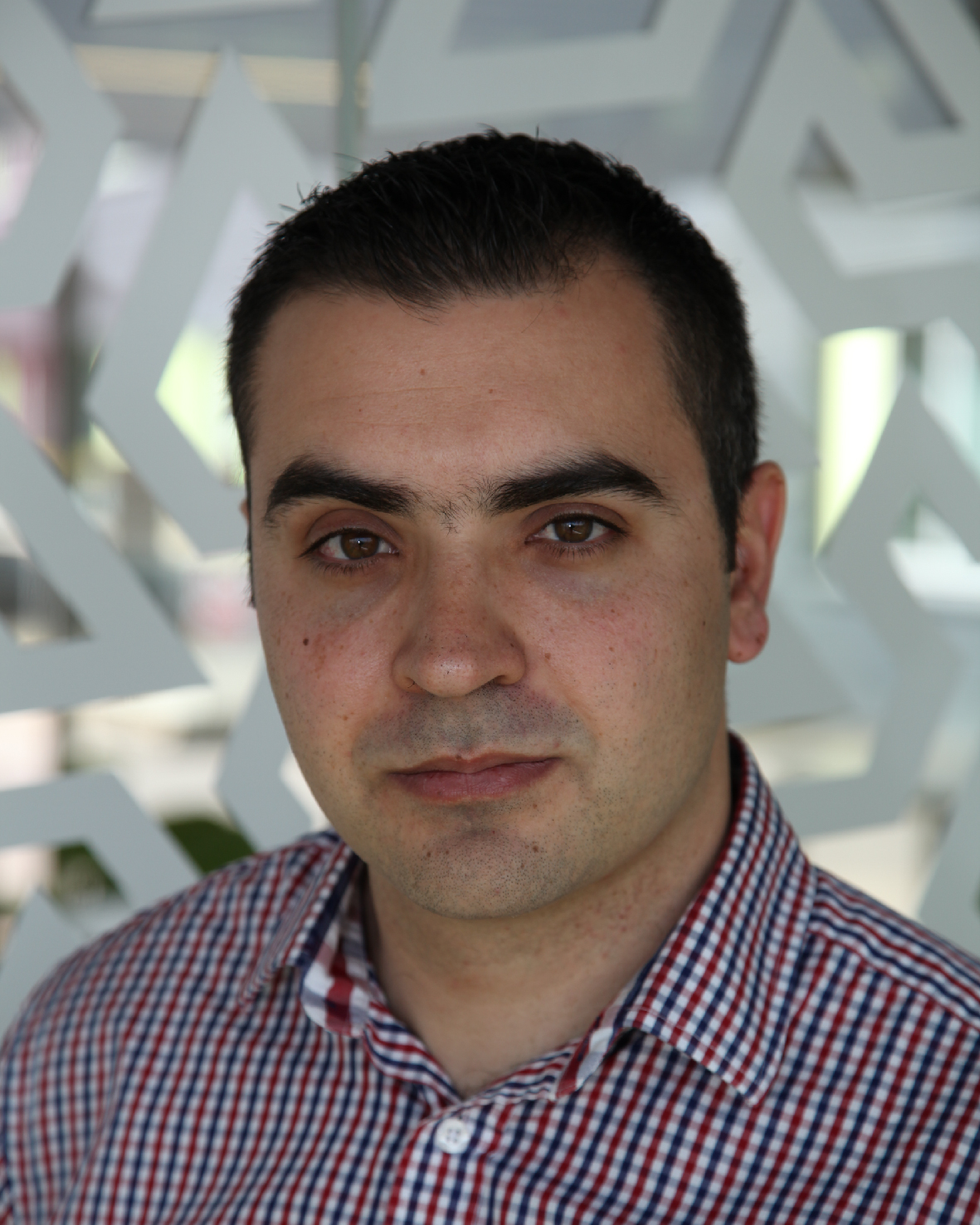}}]{Themistoklis Prodromakis}
(M'08) is a Reader in Nanoelectronics and EPSRC Fellow at the Nano Group and the Southampton Nanofabrication Centre of ECS at University of Southampton. His research interests are on bio-inspired devices for biomedical applications.
\end{IEEEbiography}

\vfill

\end{document}